# COULD ADRENALINE POSITIVE HELP TO PREVENT PSYCHOSOMATIC ILLNESSES ?

J. P. de Oliveira Filho
Projeto Phoenix
Avenida Duque de Caxias, 1448/Casa 49 – Belém, Pará, Brasil
M. Cattani
Instituto de Física da USP, 05389-970, São Paulo, SP
mcattani@if.usp.br
J. M. F. Bassalo
Faculdade de Física da UFPA and Fundação Minerva – Belém, Pará, Brasil
www.bassalo.com.br
N. P. C. de Souza
Escola de Aplicação da UFPA – Belém, Pará, Brasil
npcoelho@yahoo.com

ABSTRACT: Although the existence of a cause and effect relationship between emotions and Psychosomatic Illnesses (manifestation of organic diseases produced by emotional problems) is an unquestionable fact, yet it remains to be explained the mechanism by which an emotion affects the cure or worsening of an illness. In this paper we will examine some articles written by doctors and psychologists which confirm the Prevention Effect (or even cure) of such illnesses, and hypothesize that this cause and effect relationship is mediated by a substance, D (+) - adrenaline, which is one of the enantiomers [1] of the adrenaline molecule.

1. Introduction

The North American doctor Dean Ornish in one of his books [2] presents a series of stories based on more than twenty years of his professional experience, in which he observes that *"our survival depends on the healing power of love, intimacy, and relationships"*. [3]

In turn, the North American teacher and thinker Louise L. Hay after a personal experience (the cure of a cancer), started to study the patterns which generates physical illnesses. Thus, by means of conferences, seminaries and training programs, she started to help sick people to develop an understanding of their illnesses and to seek the cure. The success of this activity moved her to write, from 1976 on, a series of books, among which we detach two of them[4], in which she shows the relationship between illnesses and its probable emotional cause. She also suggests a change in the emotional posture of the sick person, basically a new pattern of thought (Mind), which will help the healing process (Body).

The Brazilian psychologist, psychoanalyst and physicist Maria Beatriz Breves Ramos has been developing a project similar to the two previously cited. In two of her books [5] she presents a thesis that Human Beings is a Macromicro and that he will get sick when a disequilibrium manifests itself on a biological and psychological level.[6] In a recent book, [7] the researcher Maria Beatriz shows the evidence of her thesis, by following up several patients interned in a General-Hospital, with a variety of illnesses.

In order to provide for the continuity of the arguments of our hypothesis, let us see some information about Adrenaline ($C_9H_{13}NO_3$), [8] also known as epinephrine, gotten in the research carried through the Internet by two of the authors (JMFB and NPCS, with the contribution of the Brazilian Biophysicist Alan Wilter Sousa da Silva).

It is a hormone secreted by the suprarenal glands (located above the kidneys). Adrenaline is released in little "packets". Each cell that make up the gland medulla has about 30,000 packets containing this hormone [9]. The adrenaline molecule has two enantiomers: L (-) - adrenalin and D (+) – adrenalin. The first one is ten (10) times more powerful than the second. [10] It has been well known since a long time, the effect of the L (-) - adrenaline on human beings. For example, in the face of extreme danger, stress or other life-threatening experience, adrenaline is released into the bloodstream, which provokes an increase in heart rate, a rise in the blood sugar level, a reduction of the flow of blood through the vessels and in the intestine, while maximizes this flow towards the voluntary muscles in the legs and arms and 'burns' fat in the adipose cells.

Moreover, the advancement of medical research, has revealed new uses for L(-) adrenaline such as: in local anesthesia, [11] in medical emergencies, [12] and in patients with rheumatoid arthritis. On the latter, however, that enantiomer of adrenalin was found to produce both Pro and Anti inflammatory effects .[13]

On the other hand, since the decade of 1950, studies have been made on the effects of the two enantiomers of adrenaline on rats.[14, 15] However, e-mails exchanged between the authors and some scientists, [16] associated with intensive searches in the Internet, indicate that, so far, the D (+) - adrenaline has not been tested on human beings. Moreover, no endogenous D (+) - adrenaline has been found in the human body yet. However, it is known that the L (-) - adrenaline in solution is inactivated by racemization, that is, half of it is transformed into D (+) - adrenaline. [10, 17]

2. Hypothesis

Now, let us see the hypothesis presented in this article. The Healing Effect of emotion, probably by means of a Biochemical action, was also experienced by one of the authors (JPOF). He suffered from gastric hyperacidity which tormented him. On a certain day, after some parachute jumps, he perceived that he was cured of the gastric hyperacidity. Being a Psychiatrist, with more than twenty years of experience, he made the obvious question: What had happened? He then started to reflect on what had happened and intuitively perceived that such result had been a consequence of the feelings he experienced when practicing parachute jumping.

Therefore, for him, the parachute jump experience, reproduced in every sense some techniques that he had studied in his life as a Medical Therapist, such as the Tibetan Buddhist psychology, specifically the technique of the death and rebirth of the ego. [18] JOPF also established a relationship between what he had experienced during the jump and his readings of the rituals of the passage of the Xamanic cults. In sight of these reflections, JPOF concluded that in the parachute jumping, one has exactly the same experience. In this manner, when a parachutist jumps from an airplane, it is as if he is jumping to death. It doesn't matter how much safe and secure he knows this sport is. His unconscious still doesn't know it. JPOF observes that during the jump the parachutist is living the myth of Ícarus which translates into our ancestral desire to fly and finally reborn from the ashes as the Myth of Phoenix.

The indescribable sensation at each jump (confirmed in conversations with several parachutists who had felt that parachute jumping had transformed their lives), led JPOF to hypothesize that each jump must produce biochemical reactions which generates that sensation. [19] In view of this, he started to apply this technique, with excellent results, to some of his patients with psychological upheavals, such as: pathological shyness, traumatic grief, some phobias, reactive depression, anxiety and chemical dependency. In sight of the good results obtained, it occurred to him that such results were due to the fact that these patients had experienced emotions never felt before.

Then, JPOF started to wonder what was really happening with the secretion produced by the Adrenal Glands just before the jump? Surely what was being released was not endomorphines. He realized that probably another form of adrenaline, which he denominated "Sweet Adrenaline", was involved. According to his reasoning, the "Sweet Adrenaline" would cancel (racemization) the effect of the known form of adrenaline ("Bitter Adrenaline" according to him) provided the patients could understand and control their emotions. And, by doing so, he could prevent the psychosomatic Illnesses. With this idea in mind, he spoke with one of the authors of this paper (JMFB) who brought to his attention the existence of enantiomers, pointing out the famous Thalidomide case [20]. From then on, the authors begun to develop the following Hypothesis:

*When a person experiences an emotional shock, L(-)Adrenaline is released into the bloodstream. And in its flow, the molecules start to interact, with some of them changing into D(+)-adrenaline. Under normal emotional conditions the percentage of D(+)-adrenaline in vivo remains low, such as what happens when in vitro. However, depending on the emotional control of the person,* **there might happen an increase** *on that percentage, which in turn, might cause the inactivation of the L (-) due to the racemization, a process by which half of L (-) is transformed into D (+). In this manner, this inactivation will be able to prevent a psychosomatic illness caused solely by the action of the (-) enantiomer.*

# Aknowledgments

We wish to express our grateful acknowledgment to the critical reading of Alan Wilter Sousa da Silva, José Perilo da Rosa Neto, Maria Beatriz Breves Ramos e Pedro Leon da Rosa Filho.

## NOTES AND REFERENCES

[1] Between 1848 and 1850, the French chemist Louis Pasteur studied crystals of racemic acid (from the Latin word racemus, which means grape), with the aid of a microscope. When Pasteur was observing these crystals, he noticed that there were two types of them, one being the mirror (specular) image of the other. On his quest, Pasteur used tweezers to carefully separate these crystals into two piles. He was surprised to noticed that, once separated, each new crystal would twist light in a certain direction, one clockwise, the other counterclockwise. He also noticed that when "I took an equal weight of each of the two kinds of crystals, the mixed solution was indifferent towards the light in consequence of the neutralization of the two equal and opposite individual deviations." So Pasteur inferred that the optical inactivity of the racemic acid was due to a 50% X 50% mixture (racemization) of the two types of crystals. Furthermore he observed that one of the two forms of the racemic acid was identical to the tartaric acid ($C_4H_6O_6$). In view of this, he classified the molecules which compose the studied crystals into two types: Left handed or Levorotatory [L(-)] and right handed or Dextrorotatory [D(+)] molecules. Today, these molecules known as Chiral (from the greek word "*keir*" which means hand), are called enantiomers. There are 2 types: The L(-) – enantiomer (L from levorotatory) and D(+) - enantiomer (D from dextrorotatory).

On the continuation of his research, as he observed the relationship between molecular asymmetry and microorganisms, Pasteur became convinced that the chemistry of life presented a preference for the chirality of certain molecules, and that therefore, there was a clear distinction between alive matter and dead matter. This certainty led him to present before the French Academy of Sciences his famous hypothesis: The Universe is Dissymmetric.

As time went on, Pasteur's conjecture proved to be true, and in the 20[th] Century, the development of Science revealed that this asymmetry of the Universe occurs in all levels, from the microscopic to the macroscopic, specially as for the chemistry of life. For more information on the enantiomeric forms of organic molecules see: BASSALO, J.M.F. and CATTANI, M.S.D. 1995. *Contactos* 10, p. 20; *Revista Brasileira de Ensino de Física* 17, p. 224.

[2] ORNISH, D. 1998. Love & Survival: The Scientific Basis for the Healing Power of Intimacy and Love, Rocco Publishing Company.

[3] This healing power is based on the thesis of the North American doctor Candace B. Pert, according to whom, there are no distinction between body and mind and, therefore, the two forms the body-mind system, in such a way that emotions are completely linked on to physiology.

[4] HAY, L.L. 2002. Heal Your Body: The Mental Causes for Physical Illness and the Metaphysical Way To Overcome Them, Hay House Publishing Company; ----------. 2002. You can Heal Your Life, Hay House Publishing Company.

[5] RAMOS, M.B.B. 1998. Macromicro: A Ciência do Sentir, Mauad Publishing Company; ----------. 2001. O Homem Além do Homem, Mauad Publishing Company.

[6] According to Ramos, when a person receives an emotional shock (enters in resonance with another person), his macromicro complex vibrates strongly which provokes a biological disequilibrium which in turn is translated into an alteration on the somatic level (illness).

[7] RAMOS, M.B.B. 2005. A Fronteira do adoecer: O Sentir e a Psicosomatica. Mauad Publishing Company. (This book had the contribution of the Brazilian psychologist Ana Helena Vieira Winter.)

[8] Adrenaline was discovered, independently, by four researchers: the North American doctor William Horatio Bates, in 1886; the Polish Physiologist Napoleon Cybulski, in 1895; the North American biochemist John Jacob Abel, in 1897; and the Japanese biochemist Jokichi Takamine, in 1901, who, by the way, came up with its name: ad (Latin prefix that means proximity), renal (relative to the kidneys, "renalis" in Latin) and ina (chemical suffix used in the naming of some substances). Adrenaline was artificially synthesized, in 1904, by the German chemist Friedrich Stolz.

[9] See http:/www.med.wayne.edu/pharm/artalejo.htm.

[10] SÄNGER-VAN DE GRIEND, C. E., EK, A. G., WIDAHL-NÄSMAN, M. E. and E. K. M. ANDERSSON, E. K. M. 2006. *Journal of Pharmaceutical and Biomedical Analysis* 41, p. 77;
http://cat.inist.fr/?aModele=afficheN&cpsidt=17619295;
htpp://www3.interscience.wiley.com/cgibin/abstract/107583907/
ABSTRACT?CRETRY=1&SRETRY=0.

[11] STEPENSKY, D., CHORNY, M., DABOUR, Z. and SCHUMACHER, I. 2003. *Journal of Pharmaceutical Sciences* 93, p. 969.

[12] LE COUTEUR, P. 2006; KAUMANN, A. 2006. Communication by e-mail.

[13] See http://rheumatology.oxfordjournal.org/cgi/content/full/41/9/1031.

[14] BERNHEIMER, H., EHRINGER, H., HEISTRACHER, P., und KRAUPP, O. 1960. *Boichemische Zeistchrift* **332**, p. 416.

[15] RICE, P. J., MILLER, D. D., SOKOLOSKI, T. D. and PATIL, P. N. 1989. *Chirality* 1, p. 14.

[16] P. Le Couter e A. Kaumann.

[17] Racemization of the enantiomers is generated by the interaction of molecules L (-) or D (+) with the molecules of the medium in which they are immersed CATTANI, M. and TOMÉ, T. 1993. *Origins of Life and Evolution of the Biosphere* 23, p. 125.

[18] This practice was so efficient that it was occidentalized by other psychological schools of thought such as the Gestalt, Transpersonal Psychology and the Spiritual Emergency. [Regarding this last one see: GROF, S. and GROF, C. 1989. Spiritual Emergency, Tarcher Publishing Company.]

[19] Other results on the organic effect in parachutists can be found in the following sites:
http://www.pnas.org/cgi/reprint/91/22/10440?maxtoshow=&HITS=10&hits=
10& RESULTFORMAT=& searchid=1& FIRSTINDEX=0&
minscore=5000&resourcetype+HWCIT;
See also http://www.springerlink.com/content/xx69127j52210621/;
See also http://www.ncbi.mlm.nih.gov/sites/entrez?cmd=Retrieve& db=PubMed& list_uids=8647351& dopt=Abstract.

[20] Thanks to the chirality of enantiomeric molecules it was possible to explain the famous case of thalidomide ($C_{13}H_{10}N_2O_4$). Let us see how: In Europe, particularly in England and Germany, between 1956 and 1963, it was observed that a drug (which contained thalidomide) prescribed to pregnant women to combat morning sickness and cough, was causing the birth of thousands of deformed children. Removed from the market, the drug started to be studied. They found that while the D-enantiomeric form of thalidomide had an anti-nausea effect, the L-enantiomeric form caused defects in the embryo. [MEIERHENRICH, U. 2008. Amino Acids and the Asymmetry of Life, Springer-Verlag Berlin Heidelberg.]

Likewise, studies have shown that the effectiveness of Penicillin-G ($C_{16}H_{18}N_2O_4S$) (Penicillin was discovered, in 1928, by the Scottish bacteriologist Sir Alexander Fleming) against bacterias resulted from the fact that bacterias use D-amino acids in the construction of its cellular walls, and penicillin contains a group of L-amino acids which interferes with the synthesis of the cellular walls of the bacteria. [BASSALO and CATTANI (1995).]